\documentclass[twocolumn,showpacs,preprintnumbers,amsmath,amssymb, showkeys]{revtex4}

\usepackage{graphicx}
\usepackage{dcolumn}
\usepackage{bm}
\usepackage{epsf}

\begin{document}


\title{Non-adiabatic electron charge pumping in coupled semiconductor quantum dots}

\author{P.\,I.\,Arseyev}
 \altaffiliation{ars@lpi.ru}
\author{N.\,S.\,Maslova}%
 \email{spm@spmlab.phys.msu.ru}
\author{V.\,N.\,Mantsevich}%
 \email{vmantsev@spmlab.phys.msu.ru}
\affiliation{%
 P.N. Lebedev Physical institute of RAS, Moscow, Russia, 119991
}%
\affiliation{%
 Moscow State University, Department of  Physics,
119991 Moscow, Russia
}%

\date{\today }
6 pages, 4 figures
\begin{abstract}
The possibility of non-adiabatic electron pumping in the system of
three coupled quantum dots attached to the leads is discussed. We
have found out that periodical changing of energy level position in
the middle quantum dot results in non zero mean tunneling current
appeared due to non-adiabatic non-equilibrium processes. The same
principle  can be used for fabrication of a new class of
semiconductor electronic devices based on non-stationary
non-equilibrium currents. As an example we propose a nanometer
quantum emitter with non-stationary inverse level occupation
achieved by electron pumping.
\end{abstract}

\pacs{73.23.-b, 73.40.Gk, 85.35.Be} \keywords{D. Electronic
transport in mesoscopic systems; D. Tunneling; D. Electronic
transport in QDs; D. Coulomb interaction; D. Quantum well devices }
\maketitle

\section{Introduction}

 Electron pumping in nanoscale structures attracts much attention nowadays
\cite{Blumenthal},\cite{Geerligs},\cite{Anderegg},\cite{Keller},
\cite{Switkes}, \cite{Kaestner}. A great deal of the previous
research works have been devoted to adiabatic electron pumping, the
idea discussed by Thouless \cite{Thouless} rather long time ago. The
first (to our knowledge) experiment on electron pumping in single
electron device was described in \cite{Geerligs}. Then experiments
in this direction were continued in a three-junction geometry by
Pothier et al. \cite{Pothier}. Two phase shifted $rf$ signals were
used to realize a single electron pump: a device with current $I=ef$
at zero bias voltage ($f$-frequency of the $rf$ signal). Adiabatic
charge pumping based on periodical variation of the potential
barriers formed by the finger gates  was also recently investigated
in \cite{Blumenthal}. In these systems quantized current is
connected with periodic adiabatic changing of the population of the
quantum dot.

Proposed in a number of papers photo-assisted tunneling through
coupled quantum dots \cite{Stoof},\cite{Covington}, \cite{Brune},
\cite{Stafford} is also an example of an electron pump. Pumping
effect is achieved by applying an oscillating signal to the gate
electrode or by irradiating the structure by monochromatic
\cite{Stafford} and pulsed \cite{Hazelzet} microwaves.

It was understood that for practical realization of quantized
electron pump the phenomenon of Coulomb blockade is very important
\cite{Cota}. General approach to the pumping through the interacting
quantum dots in this regime is based on supposition that the charge
relates to instantaneous chemical potential of a dot
\cite{Splettstoesser}. Using Coulomb blockade ideas another class of
non-adiabatic quantized pumping of electrons in hybrid normal
metal-superconductor structures was proposed in \cite{Pekola},
\cite{Averin}. These systems are more like a "turnstile" rather than
a "pump" because quantized current directly proportional to gate
frequency
 appears at finite (nonzero) value of applied bias. But at low temperatures these systems are very promising as
current standards \cite{Maisi}.

Adiabatic charge pumping through three tunnel-coupled quantum dots
attached to electron leads in the regime of strong Coulomb blockade
was investigated in \cite{Rezoni}. Slow variations of coupling
strength between the dots lead to adiabatic changes of energy levels
in the system. Time dependent charge redistribution caused by energy
levels changes results in non-stationary adiabatic tunneling
current.

In the present paper we suggest a new type of electron pumping also
in a system with three quantum dots, but based on non-equilibrium
non-stationary tunneling currents.
 The proposed device requires only a
single ac gate signal contrary to other semiconductor devices which
require at least two $rf$ signals with a definite phase shift. Mean
current appears in our model due to non-adiabatic changing of
electron level in a single quantum dot.

\section{Three dot model of electron pump}

\begin{figure} [t]
\includegraphics{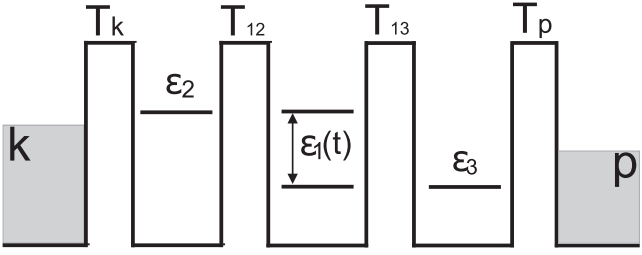}
\caption{Schematic diagram of the three coupled quantum dots with
energy level position in the middle quantum dot depending on the
time.}\label{figure1}
\end{figure}

We investigate non-stationary currents which flow in a three dot
system shown in Fig.\ref{figure1}. The left and right dots have
energy levels $\varepsilon_2$ and $\varepsilon_3$ constant in time.
And the level position of the middle dot $\varepsilon_1$ is
modulated by external gate voltage.

Quantum dots with energy levels $\varepsilon_2$ and $\varepsilon_3$
are also coupled to continuous spectrum states - massive leads.
Hamiltonian of the system under investigation has the form:
\begin{eqnarray}
\hat{H}=\sum_{i=1}^{3}\varepsilon_{i}c_{i}^{+}c_{i}+\sum_{k}\varepsilon_{k}c_{k}^{+}c_{k}+\nonumber\\
+T_{12}(c_{1}^{+}c_{2}+c_{2}^{+}c_{1})+T_{13}(c_{1}^{+}c_{3}+c_{3}^{+}c_{1})+\nonumber\\
+\sum_{k}T_{k}(c_{k}^{+}c_{2}+c_{2}^{+}c_{k})+\sum_{p}T_{p}(c_{p}^{+}c_{3}+c_{3}^{+}c_{p})
\end{eqnarray}

$T_{12}$, $T_{13}$ are tunneling transfer amplitudes between the
quantum dots and amplitudes $T_{k}$ and $T_{p}$ correspond to the
tunneling processes between the quantum dots and continuous spectrum
states. $c_{i}^{+}/c_{i}$ and $c_{k(p)}^{+}/c_{k(p)}$- electrons
creation/annihilation operators in the quantum dots localized states
and in the continuous spectrum states correspondingly. Energy values
satisfy the following ratios: $\varepsilon_2>\varepsilon_F$ and
$\varepsilon_3<\varepsilon_F$. Pumping effect appears if gate
voltage switches the level  $\varepsilon_1$ between energies
$\varepsilon_2$ and $\varepsilon_3$.

In order to develop a theory for non-stationary current let us first
describe in details
 charge relaxation processes in the system if
we assume that at the initial moment all charge density in the
system is localized in the first (middle) quantum dot and has the
value $n_{1}(0)$. At the first step we need to calculate exact
retarded Green functions of the system. In the absence of tunneling
between the three quantum dots Green functions
$G_{11}^{R}(t-t^{'})$, $G_{22}^{R}(t-t^{'})$ and
$G_{33}^{R}(t-t^{'})$ are equal to:
\begin{eqnarray}
G_{11}^{R}(t-t^{'})&=&-i\Theta(t-t^{'})e^{-i\varepsilon_1(t-t^{'})}\nonumber\\
G_{22}^{R}(t-t^{'})&=&-i\Theta(t-t^{'})e^{-i\varepsilon_2(t-t^{'})-\gamma_2(t-t^{'})}\nonumber\\
G_{33}^{R}(t-t^{'})&=&-i\Theta(t-t^{'})e^{-i\varepsilon_3(t-t^{'})-\gamma_3(t-t^{'})}
\end{eqnarray}

where $\gamma_2=\pi\nu_{k}^{0}T_{k}^{2}$ and
$\gamma_3=\pi\nu_{p}^{0}T_{p}^{2}$ are tunneling relaxation rates
from leftmost and rightmost dots respectively to the leads. Exact
retarded electron Green's function $G_{11}^{R}$ in the first quantum
dot can be found from the integral equation:

\begin{eqnarray}
G_{11}^{R}=G_{11}^{0R}+G_{11}^{0R}T_{12}^{2}G_{22}^{R}G_{11}^{R}+G_{11}^{0R}T_{13}^{2}G_{33}^{R}G_{11}^{R}
\end{eqnarray}

Acting in turn by inverse operators
$G_{11}^{0R-1},G_{22}^{R-1},G_{33}^{R-1}$ this integral equation can
be also presented in the equivalent differential form (except for
the point $t=t'$):

\begin{eqnarray}
          \label{GR}
\big[(i\frac{\partial}{\partial
t}-\varepsilon_3+i\gamma_3)((i\frac{\partial}{\partial
t}-\varepsilon_2+i\gamma_2)(i\frac{\partial}{\partial
t}-\varepsilon_1)
-T_{12}^{2}\cdot\nonumber\\
\cdot(i\frac{\partial}{\partial
t}-\varepsilon_3+i\gamma_3)-T_{13}^{2}(i\frac{\partial}{\partial
t}-\varepsilon_2+i\gamma_2)\big]G_{11}^{R}(t,t^{'})=0          \nonumber\\
\end{eqnarray}

Consequently, retarded Green's function which determine spectrum
re-normalization due to the tunneling between the quantum dots can
be written in the following form:

\begin{eqnarray}
G_{11}^{R}(t,t^{'})&=&i\Theta(t-t^{'})(A_1e^{-iE_{1}(t-t^{'})}+A_2e^{-iE_{2}(t-t^{'})}+\nonumber\\
&+&A_3e^{-iE_{3}(t-t^{'})})
\end{eqnarray}

Where eigenfrequencies $E_{1,2,3}$ can be found from equation (see
(\ref{GR})):
\begin{eqnarray}
(E-\varepsilon_1)\cdot(E-\varepsilon_2+i\gamma_{2})\cdot(E-\varepsilon_3+i\gamma_{3})-\nonumber\\
-T_{12}^{2}\cdot(E-\varepsilon_3+i\gamma_{3})-T_{13}^{2}\cdot(E-\varepsilon_2+i\gamma_{2})=0\nonumber\\
\end{eqnarray}

And coefficients $A_i$ can be evaluated using integral equation for
$G_{11}^{R}$:

\begin{eqnarray}
A_1=\frac{E_1(E_1+E_3-\widetilde{\varepsilon_2}-\widetilde{\varepsilon_3})-E_1E_3+\widetilde{\varepsilon_2}\widetilde{\varepsilon_3}}{(E_1-E_2)(E_1-E_3)}\nonumber\\
A_2=\frac{E_2(E_2+E_3-\widetilde{\varepsilon_2}-\widetilde{\varepsilon_3})-E_2E_3+\widetilde{\varepsilon_2}\widetilde{\varepsilon_3}}{(E_2-E_3)(E_2-E_1)}\nonumber\\
A_3=\frac{E_3(E_2+E_3-\widetilde{\varepsilon_2}-\widetilde{\varepsilon_3})-E_2E_3+\widetilde{\varepsilon_2}\widetilde{\varepsilon_3}}{(E_3-E_1)(E_3-E_2)}\nonumber\\
\end{eqnarray}

where $\widetilde{\varepsilon_i}=\varepsilon_i-i\gamma_i$

Further on we  assume for simplicity that $T_{12}=T_{13}=T$. If
$\varepsilon_1=\varepsilon_2$ and $\varepsilon_1-\varepsilon_3\gg
T,\gamma$ coefficients $A_i$ has the form:

\begin{eqnarray}
A_1&=&A_{1}^{0}(1+\frac{E_3-\widetilde{\varepsilon_3}}{E_1-E_3})=A_{1}^{0}(1+\frac{T^2}{(\varepsilon_1-\varepsilon_3)^2})\nonumber\\
A_2&=&A_{2}^{0}(1+\frac{E_3-\widetilde{\varepsilon_3}}{E_1-E_3})=A_{2}^{0}(1+\frac{T^2}{(\varepsilon_1-\varepsilon_3)^2})\nonumber\\
A_3&=&-\frac{T^2}{(\varepsilon_1-\varepsilon_3)^2}
\end{eqnarray}

where $A_{1}^{0}=\frac{E_1-\widetilde{\varepsilon_2}}{E_1-E_2}$
$A_{2}^{0}=-\frac{E_2-\widetilde{\varepsilon_2}}{E_1-E_2}$.

If we disconnect the third quantum dot then coefficients $A_{1}^{0}$
and $A_{2}^{0}$ give an exact solution for a system of two coupled
quantum dots for all energy values $\varepsilon_1$ and
$\varepsilon_2$.

Time evolution of electron density in the middle dot is determined
by the Keldysh Green function $G^{<}$ \cite{Keldysh}:

\begin{eqnarray}
G_{11}^{<}(t,t^{'})=in_{1}(t)
\end{eqnarray}

Equations for the Green functions $G_{ii}^{<}$ has the form:

\begin{eqnarray}
(G_{11}^{0-1}-T_{12}^{2}G_{22}^{R}-T_{13}^{2}G_{33}^{R})G_{11}^{<}=\nonumber\\=T_{12}^{2}G_{22}^{<}G_{11}^{A}
+T_{13}^{2}G_{33}^{<}G_{11}^{A}\nonumber\\
(G_{22}^{0-1}-T_{12}^{2}G_{11}^{R}-\sum_{k}T_{k}^{2}G_{kk}^{R})G_{22}^{<}=\nonumber\\=T_{12}^{2}G_{11}^{<}G_{22}^{A}
+\sum_{k}T_{k}^{2}G_{kk}^{<}G_{22}^{A}\nonumber\\
(G_{33}^{0-1}-T_{13}^{2}G_{11}^{R}-\sum_{p}T_{p}^{2}G_{pp}^{R})G_{33}^{<}=\nonumber\\=T_{13}^{2}G_{11}^{<}G_{33}^{A}
+\sum_{p}T_{p}^{2}G_{pp}^{<}G_{33}^{A}\nonumber\\
 \label{equation}
\end{eqnarray}

If $G_{22}^{<}(0,0)=i n_F(\varepsilon_2)\simeq 0$, $G_{33}^{<}(0,0)=
i n_F(\varepsilon_3)\simeq 1$ and $G_{11}^{<}(0,0)=n_1(0)$ then
Green function $G_{11}^{<}(t,t)$ is determined by the sum of
homogeneous and inhomogeneous solutions. Inhomogeneous solution of
the equation can be written in the following way:

\begin{eqnarray}
G_{11}^{<}(t,t^{'})=i T_{13}^{2}\int_{0}^{t}dt_{1}\int_{0}^{t^{'}}dt_{2}G_{11}^{R}(t-t_1)\times\nonumber\\
\times
n_F(\varepsilon_3)e^{-i\widetilde{\varepsilon_3}(t_1-t_2)}G_{11}^{A}(t_2-t^{'})    \nonumber\\
\end{eqnarray}

Homogeneous solution of the differential equation has the form:

\begin{eqnarray}
-iG_{11}^{<}(t,t^{'})=f_{1}(t^{'})e^{-iE_{1}t}+f_{2}(t^{'})e^{-iE_{2}t}+f_{3}(t^{'})e^{-iE_{3}t}\nonumber\\
\end{eqnarray}

Function $G^{<}(t,t^{'})$ satisfies the symmetry relations:

\begin{eqnarray}
[G_{11}^{<}(t,t^{'})]^{*}=-G_{11}^{<}(t^{'},t)
\end{eqnarray}

Then coefficients $f_i(t^{'})$ can be written as:
\begin{eqnarray}
f_1(t^{'})=Ae^{iE_{1}^{*}t^{'}}+Be^{iE_{2}^{*}t^{'}}+Xe^{iE_{3}^{*}t^{'}}\nonumber\\
f_2(t^{'})=Ce^{iE_{2}^{*}t^{'}}+B^{*}e^{iE_{1}^{*}t^{'}}+De^{iE_{3}^{*}t^{'}}\nonumber\\
f_3(t^{'})=X^{*}e^{iE_{1}^{*}t^{'}}+D^{*}e^{iE_{2}^{*}t^{'}}+Ze^{iE_{3}^{*}t^{'}}\nonumber\\
\label{f_i}
\end{eqnarray}

Since the solution has to satisfy homogeneous integro-differential
equation, we are able to determine all coefficients. After some
calculations we obtain that the following proportionality takes
place:
\begin{eqnarray}
f_2(t^{'})=   F_{21}  f_1(t^{'})                   \nonumber\\
f_3(t^{'})=   F_{31}  f_1(t^{'})   \nonumber
\end{eqnarray}

with coefficients $F_{21}, F_{31} $:
\begin{eqnarray}
F_{21}
=&-&\big[(E_2-\widetilde{\varepsilon_2})(E_2-\widetilde{\varepsilon_3})((E_1-\widetilde{\varepsilon_2})
(E_3-\widetilde{\varepsilon_2})+\nonumber\\
&+&(E_1-\widetilde{\varepsilon_3})(\widetilde{\varepsilon_2}-E_3))\big]\cdot\nonumber\\
&\cdot&\big[(E_1-\widetilde{\varepsilon_2})(E_1-\widetilde{\varepsilon_3})((E_3-\widetilde{\varepsilon_3})(E_2-\widetilde{\varepsilon_2})+\nonumber\\
&+&(\widetilde{\varepsilon_2}-E_3)(E_2-\widetilde{\varepsilon_2}))\big]^{-1}\nonumber\\
F_{31}
&=&\frac{(E_3-\widetilde{\varepsilon_2})(E_2-\widetilde{\varepsilon_2}+(E_1-\widetilde{\varepsilon_2})F_{21}
)}
{(E_2-\widetilde{\varepsilon_2})(E_1-\widetilde{\varepsilon_2})}\nonumber\\
\label{coefficients_1}
\end{eqnarray}

Now we can find all coefficients in (\ref{f_i}) :

\begin{eqnarray}
A&=&\frac{n_1(0)}{1+\big|F_{21}+F_{31}\big|^{2}+2ReF_{21}+2ReF_{31}}\nonumber\\
B &=& F_{21}^{*}\cdot A; \quad C=|F_{21}|^{2}\cdot A          \nonumber\\
D&=&F_{31}^{*}F_{21} \cdot A; \quad Z=|F_{31}|^{2} \cdot A;\quad X=F_{31}^{*} \cdot A\nonumber\\
\label{coefficients}
\end{eqnarray}

Finally, time dependence of the filling number in the middle quantum
dot $n_{1}(t)$ can be written as:
\begin{eqnarray}
n_{1}(t)=n_{1}^{0}\cdot(Ae^{-i(E_{1}-E_{1}^{*})t}+Ce^{-i(E_{2}-E_{2}^{*})t}+\nonumber\\
+Ze^{-i(E_{3}-E_{3}^{*})t})+2Re(Be^{-i(E_{1}-E_{2}^{*})t})+\nonumber\\
+2Re(Xe^{-i(E_{1}-E_{3}^{*})t})+ 2Re(De^{-i(E_{2}-E_{3}^{*})t})
\label{filling_numbers_1}
\end{eqnarray}

We see that there are six typical time scales in the considered
system, which are described by the expression
(\ref{filling_numbers_1}). Three of them we can identify as three
relaxation modes with rates $2 \rm{Im}E_{1}$, $2 \rm{Im}E_{2}$ and
$2 \rm{Im}E_{3}$ . Three other time scales are determined by the
expressions $\rm{Re}(E_{1}-E_{2}^{*})$, $\rm{Re}(E_{1}-E_{3}^{*})$
and $\rm{Re}(E_{2}-E_{3}^{*})$. These time scales are related with
charge density oscillations between quantum dots, if the following
ratio between $T_{ij}$ and $\gamma$ takes place:
$T_{ij}/\gamma>1/\sqrt{2}$.

If we neglect for a moment the tunneling from the middle dot to the
right one, then for the system of two coupled quantum dots
($T_{13}=0$) only three time scales appear and the equations
(\ref{coefficients_1}),(\ref{coefficients}),(\ref{filling_numbers_1})
are transformed in the following way:

\begin{eqnarray}
\frac{f_{1}(t^{'})}{f_{2}(t^{'})}=-\frac{\varepsilon_2-E_{1}-i\gamma}{\varepsilon_2-E_{2}-i\gamma}
\end{eqnarray}

Time dependence of the filling numbers in the first quantum dot
$n_{1}(t)$ can be written as:

\begin{eqnarray}
n_{1}(t)&=&n_{1}^{0}\cdot\left[Ae^{-i(E_{1}-E_{1}^{*})t}+2Re(Be^{-i(E_{1}-E_{2}^{*})t})+\right.\nonumber\\
&+&\left. Ce^{-i(E_{2}-E_{2}^{*})t}\right] \label{filling_numbers_2}
\end{eqnarray}

where coefficients $A$, $B$ and $C$ are determined as:

\begin{eqnarray}
A&=&\frac{|E_{2}-\varepsilon_1|^{2}}{|E_{2}-E_{1}|^{2}}\nonumber\\
C&=&\frac{|E_{1}-\varepsilon_1|^{2}}{|E_{2}-E_{1}|^{2}}\nonumber\\
B&=&-\frac{(E_{2}-\varepsilon_1)(E_{1}^{*}-\varepsilon_1)}{|E_{2}-E_{1}|^{2}}
\end{eqnarray}
and coefficients $X$, $Z$ and $D$ are equal to zero.

In this situation we can distinguish two relaxation rates
$\gamma_{res}$ and $\gamma_{nonres}$ which characterises charge
relaxation through an intermediate quantum dot in resonant and
nonresonant cases:

\begin{eqnarray}
\gamma_{res}=\frac{2T^2}{\gamma} \qquad
\gamma_{nonres}=\gamma_{res}\frac{\gamma^2}{(\varepsilon_1-\varepsilon_2)^2}
\end{eqnarray}

As we consider $\varepsilon_1-\varepsilon_3\gg T,\gamma$ then
$\gamma_{res}\gg \gamma_{nonres} $ and small parameter
$\gamma_{nonres}/ \gamma_{res}$ exists in the theory. This allows us
to write the following approximate relations for the system of three
coupled quantum dots in the case $\varepsilon_1 \simeq
\varepsilon_2$ valid in the first order of the small parameter
 $\frac{\gamma^2}{(\varepsilon_1-\varepsilon_3)^2} $

\begin{eqnarray}
E_{1}-E_{1}^{*}&=&-i\gamma_{res}\left[1+\frac{\gamma^{2}}{(\varepsilon_1-\varepsilon_3)^{2}}\right]\nonumber\\
E_{2}-E_{2}^{*}&=&-2i\gamma\left[1-\frac{T^{2}}{\gamma^{2}}
+\frac{T^{2}}{\gamma^{2}}\frac{T^{2}}{(\varepsilon_1-\varepsilon_3)^{2}}\right]\nonumber\\
E_{3}-E_{3}^{*}&=&2i\gamma\left[1-\frac{T^{2}}{(\varepsilon_1-\varepsilon_3)^{2}}\right]\\
E_{2}-E_{3}^{*}&=&\varepsilon_1-\varepsilon_3-2i\gamma\left[(1+\frac{T^{2}}{2\gamma^{2}}
-\frac{T^{2}}{2(\varepsilon_1-\varepsilon_3)^{2}}\right]\nonumber\\
E_{1}-E_{3}^{*}&=&\varepsilon_1-\varepsilon_3-i\gamma_{res}\frac{\gamma^{2}}{(\varepsilon_1-\varepsilon_3)^{2}}
-i\gamma\left[1-\frac{T^{2}}{(\varepsilon_1-\varepsilon_3)^{2}}\right]\nonumber\\
E_{1}-E_{2}^{*}&=&i\gamma+\frac{2T^{2}}{\varepsilon_1-\varepsilon_3}
-i\frac{T^2}{\gamma}\left[\frac{\gamma^{2}}{(\varepsilon_1-\varepsilon_3)^{2}}
-\frac{T^{2}}{(\varepsilon_1-\varepsilon_3)^{2}}\right]\nonumber\
\end{eqnarray}

When $\varepsilon_1=\varepsilon_2$ and
$\varepsilon_1-\varepsilon_3\gg T,\gamma$ the exact equations
(\ref{coefficients_1}) can be transformed in the following way .

\begin{eqnarray}
F_{21}\simeq
-\frac{T^{2}}{\gamma^{2}}\left[1+\frac{\gamma}{\varepsilon_1-\varepsilon_3}+\frac{T^{2}}{\gamma^{2}}-
\frac{T^{2}}{\gamma(\varepsilon_1-\varepsilon_3)}\right]\nonumber\\
F_{31} \simeq
\frac{T^{2}}{\gamma^{2}}\left[\frac{T^{2}}{(\varepsilon_1-\varepsilon_3)^{2}}+i\frac{T^{2}}
{(\varepsilon_1-\varepsilon_3)^{2}}\right]
\end{eqnarray}

So, coefficients $D$, $Z$ and $X$ which are resposible for the
"reverse" current to the right lead are much smaller than $A$, $B$
and $C$, which correspond to "direct" current to the left, due to
the appearance of the parameter
$\frac{T^{2}}{(\varepsilon_1-\varepsilon_3)^{2}}$. For simplicity we
omit the terms with coefficients $D$, $Z$ and $X$ in equation
(\ref{filling_numbers_1}) which determine time evolution of
localized charge in the middle quantum dot. For any concrete system
the accuracy of this approximation can be easily estimated from the
exact equations.

Pumping of electrons takes place if energy level $\varepsilon_1(t)$
is a function of time and changes periodically (Fig. \ref{figure1}).
We shall describe the most favorable case with $T\ll\gamma$. For
current calculation we consider the situation of periodically
switching the position of level $\varepsilon_1$ by external gate:

$\varepsilon_1(t)=\varepsilon_3$ in the interval $0<t<t_0$ it means
resonant tunneling between energy levels $\varepsilon_1$ and
$\varepsilon_3$

$\varepsilon_1(t)=\varepsilon_2$ in the interval $t_0<t<2t_0$ -
resonance between energy levels $\varepsilon_1$ and $\varepsilon_2$

\begin{figure} [h]
\includegraphics[width=50mm]{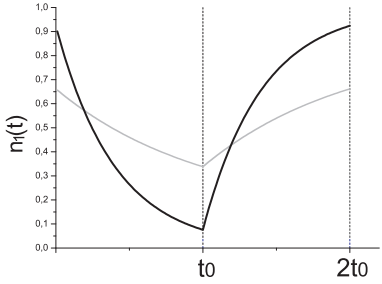} \caption{Time evolution of local
electron density $n_{1}(t)$ in the central quantum dot in the system
of three coupled quantum dots if energy level position in the middle
quantum dot depends on time. Black line corresponds to the case when
frequency of energy level $\varepsilon_{1}(t)$ switching is lower
than tunneling rates and grey line corresponds to the case when
frequency is higher than tunneling rates.}\label{figure2}
\end{figure}

Time evolution of local electron density $n_{1}(t)$ in the central
quantum dot can be determined from equation
(\ref{filling_numbers_1}) (Fig.\ref{figure2}).

When $0<t<t_0$

\begin{eqnarray}
 n_1(t)=n_1^0\left[\left(1+\frac{\gamma_{res}}{\gamma}\right)e^{
 -{\displaystyle \gamma_{res}t}} -
 \frac{\gamma_{res}}{\gamma}e^{\displaystyle-{{\gamma}}t}
 \right]\nonumber\\
\end{eqnarray}

and when $t_0<t<2t_0$

\begin{eqnarray}
n_1(t)=n_1^0\big[\big(1+\frac{\gamma_{res}}{\gamma}\big)e^{
-{\displaystyle \gamma_{res}(t-t_0)}}-\nonumber\\
-\frac{\gamma_{res}}{\gamma}(e^{\displaystyle-{{\gamma}}(t-t_0)}
\big]+\nonumber\\
+\big[1-(1+\frac{\gamma_{res}}{\gamma})e^{\displaystyle-{\gamma_{res}(t-t_0)}}+\nonumber\\
+\frac{\gamma_{res}}{\gamma} e^{\displaystyle-{{\gamma}}(t-t_0)} \big]\nonumber\\
\end{eqnarray}

Taking into account periodicity condition $n_1(2t_0)=n_1^0$, one can
find $n_1^0$:

\begin{eqnarray}
n_1^0=\frac{1}{1+
\left(1+\frac{\gamma_{res}}{\gamma}\right)e^{-{{\displaystyle
\gamma_{res}t_0}}}
-\frac{\gamma_{res}}{\gamma}e^{\displaystyle-{{\gamma}}t_0}}
\end{eqnarray}

Results for $n_1(t)$ are shown in Fig.\ref{figure2}. Situation when
frequency $\Omega\equiv 1/2t_0$ of the level $\varepsilon_{1}(t)$
switching is higher than tunneling rates $\gamma_{res}, \gamma$ is
depicted by grey line. Black line corresponds to the case when
frequency $\Omega$ is lower than $\gamma_{res}, \gamma$. It is clear
that with the increasing of frequency the value of $n_{1}(t)$ always
tends to the value $1/2$ and is almost independent on time at high
gate frequencies.

For low frequencies if $\gamma_{res}t_0\gg 1$ the value $n_1^0$ is
almost equal to 1. The energy level $\varepsilon_1$ is filled up
almost completely during the pumping cycle (for considered situation
when energy level $\varepsilon_2$ is well above and energy level
$\varepsilon_3$ is well below the Fermi energy). Non-stationary
tunneling current through the system appears for zero applied bias:
\begin{eqnarray}
e\frac{\partial}{\partial t}\, n_1(t)= I(t)
\end{eqnarray}

One can find for $0<t<t_0$
\begin{eqnarray}
I(t)=e
n_1^0\gamma_{res}\left[(1+\frac{\gamma_{res}}{\gamma})e^{-\gamma_{res}t}
-e^{-\gamma t} \right]
\end{eqnarray}

Mean tunneling current value can be found as:
\begin{eqnarray}
<I>=e\frac{1}{2t_0}\, \int\limits_0^{t_0}I(t)dt =
e \frac{1}{2t_0}n_1^0\cdot\nonumber\\
\cdot\left[1-(1+\frac{\gamma_{res}}{\gamma})e^{\displaystyle-\gamma_{res}t_0}+
\frac{\gamma_{res}}{\gamma}e^{\displaystyle-\gamma t_0}\right]
\label{brackets}
\end{eqnarray}

If $\Omega\equiv 1/2t_0\ll\gamma_{res}$ tunneling current mean value
can be written as $<I>=e \Omega$ since $n_1^0=1$ for such
frequencies. This is the regime, when the device operates like a
current standart: the current is directly proportional to the gate
frequency. This regime has exponential accuracy which is governed by
the second and third terms in the square brackets in expression
(\ref{brackets}). Note, that even for very unsuitable case if
$|\varepsilon_2-\varepsilon_3|\simeq \gamma\simeq T $ the pumping
effect still remains, and the cuurrent is proportional to the
frequency, though it's value
 is suppressed compared to the ideal relation $<I>=e \Omega$.

For high frequencies of the gate voltage in the region
$\gamma\gg\Omega\equiv 1/2t_0 \gg \gamma_{res}$ tunneling current
average value is almost independent on the frequency and equal to:
$<I>=e \gamma_{res}/4$ ($ \gamma_{res}=\frac{\displaystyle
2T^2}{\displaystyle\gamma} $). With further frequency increase
($\Omega \gg\gamma\gg \gamma_{res}$) mean current value decreases to
$ <I>=e \gamma_{res}^{2}/4\gamma $.

\begin{figure} [h]
\includegraphics[width=50mm]{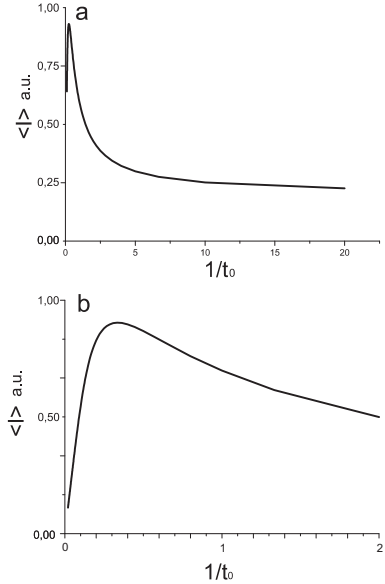} \caption{Frequency
dependence of the mean tunneling current for the system of three
coupled quantum dots (gate frequency $\Omega=1/2t_0$). Different
frequency scales are presented.}\label{figure3a_3b}
\end{figure}

 The non-stationary mean tunneling current value has non-monotonic
dependence on the gate frequency with maximum at $\Omega \simeq
\gamma_{res}$ (Fig.\ref{figure3a_3b}). This effect can be used for
frequency stabilization in nanoelectronics.

The parameters of devices based on quantum dots depend on the size
of quantum dots, tunneling transfer rates, energy levels positions
and distances between them. Estimation of tunneling parameters for
achievable setup gives us characteristic frequencies and currents
for such devices:

$T\simeq 1\,meV, \quad \gamma \simeq 1\div10 \, meV \Rightarrow$

$ \gamma_{res}=2T^2/\gamma \simeq 0.1\div 1 \, meV \simeq 10^{10}
\div 10^{11} \, 1/sec$

where parameters $T$ and $\gamma$ are determined by the widths and
heights of the barriers. For such values of the tunneling rates we
need to have quantum dots of tens of nanometers size, for which
quantum size quantization energies are
 not less than the tunneling width of levels. Such devices could operate at
gigahertz and subgigahertz frequencies at nano and subnanoampere
currents ($1nA \simeq 6 \cdot 10^{9} \, e/sec$).

\begin{figure} [h]
\includegraphics{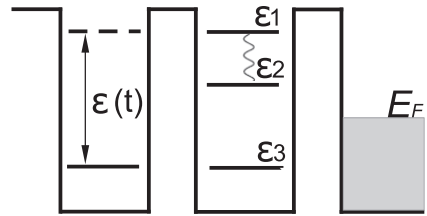} \caption{Schematic diagram of
two coupled quantum dots which operate as an emitter due to non
stationary inverse occupation of levels.}\label{figure4}
\end{figure}

The difference between tunneling rates in resonant and non resonant
cases can be used also for creation of inverse occupation in quantum
emitter based on the system of coupled quantum dots. An example of
such device is shown in Fig.\ref{figure4}.
In the system of two coupled quantum dots we switch the level
$\varepsilon$ by applying gate voltage to the second quantum dot
between two levels of the first quantum dot directly coupled to the
lead. If the third non-resonant level exists in the same quantum dot
between the lowest and the highest levels, then charge pumping from
the initially filled state $\varepsilon_3$ to the highest level
$\varepsilon_1$ creates inverse occupation of the states
$\varepsilon_2$ and $\varepsilon_1$. So some nanometer quantum
generators can be fabricated on this principle.

\section{Conclusion}

We investigated electron pumping ability of a system of three tunnel
coupled quantum dots attached to the leads. Periodical changing of
energy level position in the middle quantum dot by gate voltage
leads to nonzero tunneling current even if applied to the structure
bias is equal to zero.

Our calculations of the mean current are based on accurate analysis
of relaxation processes in quantum dots in non-adiabatic regime.
Exact equations allows to investigate various regimes of the device
and estimate the mean current value and accuracy of it's operation
as a current standard. For very small dots with pronounced size
effect a possibility of room temperature electron pumping is opened.
We should like to stress that the ideas discussed in the paper can
be used for fabrication of a new type of electronic devices based on
non-equilibrium non-stationary tunneling currents. As an example we
proposed in the paper nanometer quantum emitter based on two coupled
quantum dots.

\section{Acknowledgements}
The financial support by RFBR, Leading Scientific School grants and
grants of Russian Minestry of Science is acknowledged.


\pagebreak

\end{document}